\documentclass[%
 preprint,
showpacs,preprintnumbers,
 amsmath,amssymb,
 aps,
]{revtex4-1}
\usepackage[pdftex]{graphicx,color}
\usepackage{subfigure,
mathrsfs,braket,amsmath,amssymb,dcolumn,bm,comment}
\newlength{\subfigwidth}
\newlength{\subfigcolsep}
\setkeys{Gin}{width=\subfigwidth}
\makeatletter 
 
\@addtoreset{figure}{section} 
\makeatother 
\begin{document}
\preprint{HUPD1503}
\def\tbr{\textcolor{red}}
\def\tcr{\textcolor{red}}
\def\ov{\overline}
\def\dprime{{\prime \prime}}
\def\nn{\nonumber}
\def\f{\frac}
\def\p{\partial}
\def\H{\mathcal{H}}
\def\beq{\begin{equation}}
\def\eeq{\end{equation}}
\def\bea{\begin{eqnarray}}
\def\eea{\end{eqnarray}}
\def\bsub{\begin{subequations}}
\def\esub{\end{subequations}}
\def\dc{\stackrel{\leftrightarrow}{\partial}}
\def\d{\partial}
\def\sla#1{\rlap/#1}
\def\mH{\mathscr{H}}
\def\tD{\tilde{D}}
\def\Q{{\cal Q}}
\def\mpim{m_{\pi^-}}
\def\mpi0{m_{\pi^0}}
\def\meta{m_\eta}
\def\tauepp{\tau^- \to \nu_\tau \eta \pi^- \pi^0}
\def\Vcurrent{\bar{d}\gamma_\mu u}
\def\Acurrent{\bar{d} \gamma_\mu \gamma_5 u}
\def\VmA{\bar{d}\gamma_\mu(1-\gamma_5)u}
\def\epp{\eta \pi^- \pi^0}
\def\hg{{G}_{2p}}
\def\IP{\mathrm{IP}}
\def\hc{\mathrm{h}.\mathrm{c}.}
\def\nn{\nonumber}
\def\beq{\begin{equation}}
\def\eeq{\end{equation}}
\def\bei{\begin{itemize}}
\def\eei{\end{itemize}}
\def\bea{\begin{eqnarray}}
\def\eea{\end{eqnarray}}
\def\s{\partial \hspace{-.47em}/}
\def\ad{\overleftrightarrow{\partial}}
\def\para{%
\setlength{\unitlength}{1pt}%
\thinlines %
\begin{picture}(12, 12)%
\put(0,0){/}
\put(2,0){/}
\end{picture}%
}%
\title{Precise discussion on
T-asymmetry
with B-meson decays}


\author{Hiroyuki Umeeda}
\email[Speaker, E-mail: ]{umeeda@theo.phys.sci.hiroshima-u.ac.jp}
\affiliation{Graduate School of Science, Hiroshima University,
Core of Research for the Energetic Universe, Hiroshima University,\\
Higashi-Hiroshima, 739-8526, Japan}
\author{Takuya Morozumi}
\email[E-mail: ]{morozumi@hiroshima-u.ac.jp}
\affiliation{Graduate School of Science, Hiroshima University,
Core of Research for the Energetic Universe, Hiroshima University,\\
Higashi-Hiroshima, 739-8526, Japan}
\author{Hideaki Okane}
\email[E-mail: ]{hideaki-ookane@hiroshima-u.ac.jp}
\affiliation{Graduate School of Science, Hiroshima University,
Core of Research for the Energetic Universe, Hiroshima University,\\
Higashi-Hiroshima, 739-8526, Japan}
\begin{abstract}
Through $B\bar{B}$ mixing system,
one can construct an asymmetry which naively seems 
to be a time reversal (T) odd quantity.
In this article, the two processes $(\mathrm{a})\: B_-\rightarrow 
\bar{B^0}$ and $(\mathrm{b})\: \bar{B^0}\rightarrow B_-$ are
used to construct the event number asymmetry.
The CP violation of Kaon system denoted as $\epsilon_K$ contributes to observables and we evaluate the contribution from $\epsilon_K$ explicitly.
The asymmetry is formulated with phase convention independent parameters which are invariant under
redefinition of phase of quark fields.
The overall factors of the time dependent decay rates
are taken into account in this article.
Furthermore, we suggest conditions for the asymmetry
to be a T-odd quantity.
The one of such conditions arises due to
the difference of overall factors which form the asymmetry.
%
\end{abstract}
\maketitle
\section{Introduction}
Time reversal (T) is a fundamental symmetry in particle physics and observation of T-asymmetry plays a crucial role
to investigate the property of theoretical aspects. BaBar collaboration announced their result\cite{Lees:2012uka}
that they observed the T-asymmetry through B-meson
mixing system.
The idea of taking T-asymmetry is based on \cite{Banuls:1999aj}-\cite{Bernabeu:2012ab}.
BaBar stated that the observation is the first evidence of
T-violation in B-system, and
BaBar observable naively seems to be T-asymmetry.
However, discussion in Ref.\cite{Applebaum:2013wxa} shows that BaBar observable slightly deviates from a T-odd quantity.
\par
In this article, we construct an asymmetry which consists of
$B_d$ meson system.
The difference between BaBar asymmetry
and the one in this article is stated.
The mixing induced CP violation of Kaon system contributes to the asymmetry
so that one needs to take account of such mixing effect.
The contribution from $\epsilon_K$ to an observable is extracted.
The formulation in this article is
independent from phase redefinition of quark fields.
For further detail, one can refer the paper\cite{Morozumi:2014xwa}.
\section{Time dependent asymmetry with $B_d$-decays}
\begin{figure}[h]
\includegraphics[width=.5\textwidth]{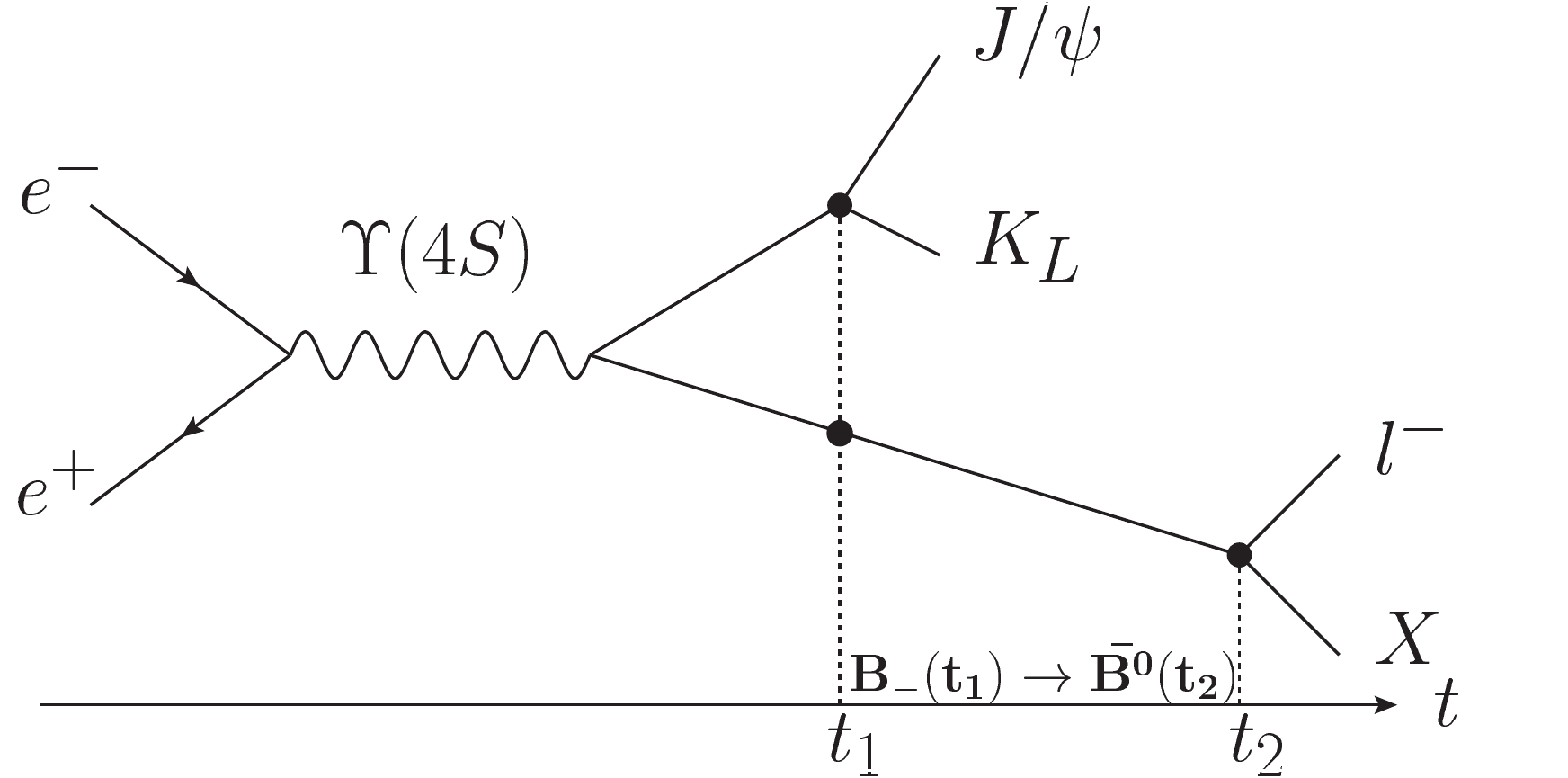}
\includegraphics[width=.5\textwidth]{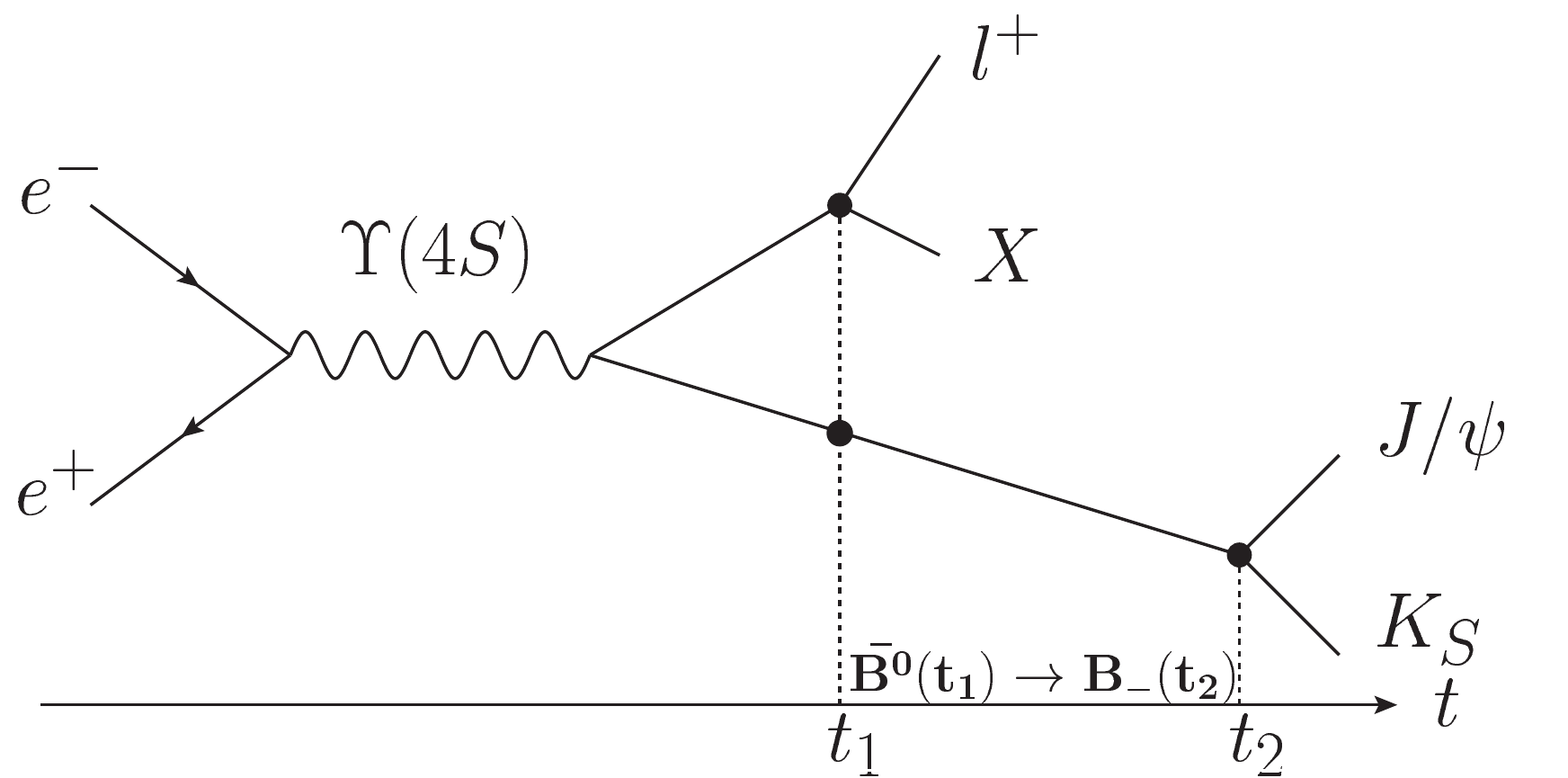}
\caption{
(Upper) Process (a) in Eq.(2.1),$\qquad\qquad$
(Lower) Process (b) in Eq.(2.1)}
\label{fig1}
\end{figure}

In this section, an event number asymmetry is constructed
and one can check how it behaves under T-transformation.
Using processes which come from $\Upsilon(4S)$ decay,
the following processes in $B_d$ system are considered,
\begin{eqnarray}
(\mathrm{a})\: B_-\rightarrow \bar{B^0},\quad
(\mathrm{b})\: \bar{B^0}\rightarrow B_-,
\label{2.1}
\end{eqnarray}
where $B_-$ indicates the CP odd state of $B_d$ meson.
The processes (a) and (b) are naively related with time reversal
since the time direction is opposite to each other.
In Fig.\ref{fig1}, the entire processes of (a) and (b)
are shown respectively.
\par
Hereafter, we explain how B mesons in (a) are identified.
In (a), CP tagging is conducted at time $t_1$ to determine $B_+$ (CP even B meson), and one also identify $B_-$
due to Einstein Podolsky Rosen (EPR) correlation from $\Upsilon(4S)$ decay.
At time $t_2$, B meson decays into $l^- X$ and the flavor tagging
method enables us to determine $\bar{B^0}$.
Hence, one can select the event of $B_-\rightarrow \bar{B^0}$.
Similarly, one can obtain events of (b)
with flavor tagging and CP tagging.\par
Time dependent asymmetry is defined as,
\begin{equation}
A(t)=
\frac{\Gamma_{(a)}(t)-\Gamma_{(b)}(t)}
{\Gamma_{(a)}(t)+\Gamma_{(b)}(t)},\label{2.2}
\end{equation}
where $\Gamma_{(a)}$ and $\Gamma_{(b)}$
are time dependent decay rates for the processes (a) and (b)
in Eq.(\ref{2.1}).
Hence, the asymmetry above is naively thought to be T-odd
quantity.
Here, time dependent decay rate is given as,
\begin{eqnarray}
&\Gamma_{(a)}=e^{-\Gamma(t_1+t_2)}N_{(a)}\kappa_{(a)}\left[
\cosh(y\Gamma t)+\displaystyle\frac{\sigma_{(a)}}{\kappa_{(a)}}\sinh(y\Gamma t)+
\displaystyle\frac{C_{(a)}}{\kappa_{(a)}}\cos(x\Gamma t)
+\displaystyle\frac{S_{(a)}}{\kappa_{(a)}}\sin(x\Gamma t)
\right],&\label{2.3}\\
&t=t_2-t_1,\quad
\Gamma=\displaystyle\frac{\Gamma_H+\Gamma_L}{2},\quad
x=\displaystyle\frac{m_H-m_L}{\Gamma},\quad
y=\displaystyle\frac{\Gamma_H-\Gamma_L}{2\Gamma},&\nonumber
\end{eqnarray}
where the coefficients $N_{(a)}, \kappa_{(a)}, \sigma_{(a)}, C_{(a)}$
and $S_{(a)}$ are given in
\cite{Applebaum:2013wxa}\cite{Morozumi:2014xwa}
and we do not specify the expression of these coefficients in this article.
Here the following quantities are introduced,
\begin{eqnarray}
&N_R=\displaystyle\frac{N_{(b)}\kappa_{(b)}}{N_{(a)}\kappa_{(a)}}
\simeq 1+\Delta N_R\label{2.5},&\\
&\Delta X=\displaystyle\frac{1}{\sqrt{N_R}}\displaystyle\frac{X_{(a)}}{\kappa_{(a)}}-
\sqrt{N_R\displaystyle}\frac{X_{(b)}}{\kappa_{(b)}},\quad
\hat{X}=\displaystyle\frac{1}{\sqrt{N_R}}\displaystyle\frac{X_{(a)}}{\kappa_{(a)}}+
\sqrt{N_R}\displaystyle\frac{X_{(b)}}{\kappa_{(b)}},\quad
(X=\sigma,\: S \:\mathrm{and}\:  C)&
\end{eqnarray}
where $N_R$ in Eq.(\ref{2.5}) is ratio of overall factors 
for the time dependent decay rates in Eq.(\ref{2.3}).
$\Delta N_R$ shows the effect of non-zero difference
of the overall factors in Eq.(\ref{2.3}).
$\Delta N_R$ consists of small parameters such like mixing induced CP violation of $B_d$ system and CP violation in $B_d\rightarrow l^-X$ decay
and $\Delta N_R$ is evaluated in Ref.\cite{Morozumi:2014xwa}.
BaBar collaboration measured the processes: 
$B_-\rightarrow\bar{B^0}$ and $\bar{B^0}\rightarrow B_-$.
BaBar-constructed asymmetry eliminates the effect of overall factors
so that the definition of BaBar asymmetry is not identical to
the event number asymmetry in Eq.(\ref{2.2}).
In Ref.\cite{Applebaum:2013wxa},
BaBar asymmetry\cite{Lees:2012uka} is analyzed.\par
In this article, we investigate the event number asymmetry
so that the contribution from $\Delta N_R$ in Eq.(\ref{2.5})
needs to be taken account and the contribution from $\epsilon_K$
is calculated.
With the time dependent decay rate in Eq.(\ref{2.3}), the asymmetry in Eq.(\ref{2.2}) is given as,
\begin{eqnarray}
A(t)\simeq\displaystyle\frac{-\displaystyle\frac{\Delta N_R}{2}+\displaystyle\frac{\Delta \sigma}{2}y\Gamma t+\displaystyle\frac{\Delta S}{2}\sin(x\Gamma t)
+\displaystyle\frac{\Delta C}{2}\cos(x\Gamma t)
}{1+\displaystyle\frac{\hat{\sigma}}{2}y\Gamma t+\displaystyle\frac{\hat{S}}{2}\sin(x\Gamma t)
+\displaystyle\frac{\hat{C}}{2}\cos(x\Gamma t)}.
\label{asym}
\end{eqnarray}
The event number asymmetry in Eq.(\ref{asym}) is expanded
as,
\begin{eqnarray}
A(t)&\simeq& R_T+C_T\cos(x\Gamma t)+ S_T \sin(x\Gamma t)\nonumber\\
&&+B_T\sin^2(x\Gamma t) +
D_T\sin(x\Gamma t)\cos(x\Gamma t)
+E_T(y\Gamma t)\sin(x\Gamma t).\label{2.6}
\end{eqnarray}
Hereafter, we discuss the coefficient of $\cos(x\Gamma t)$
and the other coefficients in Eq.(\ref{2.6}) are evaluated and discussed
in Ref.\cite{Morozumi:2014xwa}.
The following quantities are introduced,
\begin{eqnarray}
\lambda=\frac{q}{p}\frac{p_K}{q_K}
\frac{\bar{A}_{\psi \bar{K^0}}}{A_{\psi K^0}}\sqrt{\frac{1+\theta_K}{1-\theta_K}},\quad
C=\frac{1-|\lambda|^2}{1+|\lambda|^2},\;
S=\frac{2\mathrm{Im}\lambda}{1+|\lambda|^2},\quad
\theta_K=\frac{A_{\psi K^0}A^{\mathrm{ID}}_{\psi K^0}-\bar{A}_{\psi \bar{K^0}}\bar{A}^\mathrm{ID}_{\psi \bar{K^0}}}
{A_{\psi K^0}A^{\mathrm{ID}}_{\psi K^0}+\bar{A}_{\psi \bar{K^0}}\bar{A}^\mathrm{ID}_{\psi \bar{K^0}}},\label{para}
\end{eqnarray}
where $B_d$ decay amplitude is denoted $A_{\psi K^0}$ for $B^0\rightarrow \psi K^0$ and $A_{\psi \bar{K^0}}$ for
$B^0\rightarrow \psi \bar{K^0}$, while inverse decay amplitudes
are written as $A_{\psi K^0}^\mathrm{ID}$ for
$\psi K^0\rightarrow B^0$ and $A_{\psi \bar{K^0}}^\mathrm{ID}$ for
$\psi \bar{K^0}\rightarrow B^0$.
Mixing parameters in B and K systems are denoted as,
\begin{eqnarray}
&\ket{K_L}_{\mathrm{in}}=p_K\sqrt{1+z_K}\ket{K^0}-q_K\sqrt{1-z_K}\ket{\bar{K^0}},\;
\ket{K_S}_{\mathrm{in}}=p_K\sqrt{1-z_K}\ket{K^0}+q_K\sqrt{1+z_K}\ket{\bar{K^0}},\quad&\label{K}\\
&\ket{B_H}_{\mathrm{in}}=p\sqrt{1+z}\ket{B^0}-q\sqrt{1-z}\ket{\bar{B^0}},\;
\ket{B_L}_{\mathrm{in}}=p\sqrt{1-z}\ket{B^0}+q_K\sqrt{1+z}\ket{\bar{B^0}}&\label{B}
\end{eqnarray}
One can check\cite{Morozumi:2014xwa} that $C, S, \theta_K, z_K$ and $z$ in Eqs.(\ref{para}-\ref{B}) are odd or even under T-transformation.
The transformation property is exhibited in Tab.\ref{tab1}.
\begin{table}[h]
\centering
\caption{Transformation properties of the parameters
under T-transformation. These parameters are to express the asymmetry
so that one can show the transformation property of each term.
}
  \begin{tabular}{|c|c|c|c|c|c|c|}
\hline
&$C$ & $S$ & $\theta_K$ & $z_K$ & $z$&$\Delta\lambda_l$\\\hline
T & $-$& $-$&$+$&$+$&$+$&$-$\\\hline
  \end{tabular}
\label{tab1}
\end{table}\par
The coefficient $C_T$ in Eq.(\ref{2.6}) is parameterized as,
\begin{eqnarray}
C_T&=&\frac{\Delta C}{2}\simeq
C-Sz^I+\theta^R_K+S\Delta\lambda_l^I.\label{2.8}
\end{eqnarray}
In Eq.(\ref{2.8}), $C_T$ is written in terms of
parameters which transform odd or even under T-transformation as shown in Tab.\ref{tab1}.
$\Delta\lambda_l$ in Eq.(\ref{2.8}) denotes wrong sign semi-leptonic
decays of $B_d$ meson such like $B^0\rightarrow l^-X$ and $\bar{B^0}\rightarrow l^+X$.
Formulation of $\Delta\lambda_l$ is given in Ref.\cite{Morozumi:2014xwa}.
One can find that the quantities in Eq.(\ref{2.8})
include not only T-odd terms but also T-even contribution.
This slight deviation from T-odd quantities is
pointed out in Ref.\cite{Applebaum:2013wxa}
regarding to BaBar asymmetry\cite{Lees:2012uka}.
The event number asymmetry coefficients in
Eq.(\ref{2.8}) slightly deviate from
T-odd quantity as well as BaBar asymmetry.
There exists contribution from mixing induced CP violation in Kaon system since the final states in Fig.\ref{fig1} include
$K_L$ and $K_S$ which are different from exact CP eigenstates.
Extracting $\epsilon_K$ in Eq.(\ref{2.8}), one expresses,
\begin{eqnarray}
&C_T\simeq C-Sz^I+\theta_K^R+S\Delta\lambda_l^I
\simeq (C^\prime -2\mathrm{Re}\epsilon_K)
-Sz^I+\theta_K^R+S\Delta\lambda_l^I,&\label{2.13}\\
&C^\prime=\displaystyle\frac{1-|\lambda^\prime|^2}{1+|\lambda^\prime|^2},\qquad
\lambda^\prime=\displaystyle\frac{q}{p}\displaystyle\frac{\bar{A}_{\psi \bar{K^0}}}{A_{\psi K^0}}\sqrt{\displaystyle\frac{1+\theta_K}{1-\theta_K}}.&\label{2.14}
\end{eqnarray}
Under phase redefinition of a quark such like $\bra{K^0}\rightarrow e^{-i\alpha_K}\bra{K^0}$ and
$\bra{\bar{K^0}}\rightarrow e^{+i\alpha_K}\bra{\bar{K^0}}$,
$\lambda^\prime$ transforms as $\lambda^\prime\rightarrow \lambda^\prime e^{+2i\alpha_K}$.
Therefore, $C^\prime$ in Eq.(\ref{2.14})
is shown as a phase convention independent parameter.
One also sees that $\mathrm{Re}\epsilon_K$
is phase convention independent and
the quantities in Eq.(\ref{2.13})
are phase convention independent.
We find that $Re\epsilon_K\sim\mathcal{O}(10^{-3})$
arises to contribute to $C_T$ in Eq.(\ref{2.13})
and $C_T$ becomes a quantity of $\mathcal{O}(10^{-3})$ \cite{Morozumi:2014xwa}.
\par
\section{Conditions for asymmetry to be a T-odd quantity}
In the previous section, T-even parts in Eq.(\ref{2.8}) are found and the coefficients slightly deviate from T-odd quantities.
Hereafter, we discuss conditions for the coefficients in Eq.(\ref{2.6})
to be T-odd (T-violating) quantities.\par
There are two types of such a condition. The first one is (1) B-meson state equivalence which is originally suggested
and calculated in Ref.\cite{Applebaum:2013wxa}.
This condition is regarding to B-meson states
which appear in Fig.\ref{fig1} and it requires that
B meson states should be equivalent to
the states which reveal in diagrams whose time direction is totally
opposite to Fig.\ref{fig1}. (therefore such diagram contains
$e^+ e^-$ as a final state and $\psi K_L l^-X$ as an initial 
state) This is diagrammatically discussed in Ref.\cite{Morozumi:2014xwa}.
\par
The second type of the conditions is (2) $\Delta N_R^e=0$; T-even part of the ratio of the overall normalizations in Eq.(\ref{2.5}) should vanish.
Since the event number asymmetry in Eq.(\ref{2.2}) is evaluated
with non-zero $\Delta N_R$,
$\Delta N_R$ can give rise to T-even effect.
This is not the case for BaBar asymmetry\cite{Lees:2012uka}
since it eliminates the effect of overall normalization difference between the two processes (a) and (b) in Eq.(2.1).
However, when $\Delta N_R^e=0$ is satisfied, $\Delta X (X=\sigma, C, S)$ is a T-odd quantity while 
$\hat{X} (X=\sigma, C, S)$ is proven\cite{Morozumi:2014xwa} to be a T-even quantity.
Therefore, Eq.(\ref{asym}) reads as,
\begin{equation}
A(t)\simeq\displaystyle\frac{-\displaystyle\frac{\Delta N_R}{2}+\displaystyle\frac{\Delta \sigma}{2}y\Gamma t+\displaystyle\frac{\Delta S}{2}\sin(x\Gamma t)
+\displaystyle\frac{\Delta C}{2}\cos(x\Gamma t)
}{1+\displaystyle\frac{\hat{\sigma}}{2}y\Gamma t+\displaystyle\frac{\hat{S}}{2}\sin(x\Gamma t)
+\displaystyle\frac{\hat{C}}{2}\cos(x\Gamma t)}
\sim\frac{\mathrm{T}-\mathrm{odd}}{\mathrm{T}-\mathrm{even}},
\end{equation}
and A(t) turns out to be T-odd quantity. 
One can show that the event number asymmetry coefficients
in Eq.(\ref{2.6}) become T-odd quantities
when the two types
of the conditions are simultaneously satisfied\cite{Morozumi:2014xwa}.
\section{Summary}
In this article, the event number asymmetry 
is investigated for
the two processes: $(\mathrm{a})\: \bar{B^0}\rightarrow B_-$ and
$(\mathrm{b})\: B_-\rightarrow \bar{B^0}$.
The constructed asymmetry in Eq.(\ref{2.2}) appears to
include not only T-odd part but also T-even part,
as shown in Ref.\cite{Applebaum:2013wxa} with respect
to BaBar asymmetry\cite{Lees:2012uka}.
The mixing induced CP violation of Kaon system, i.e.,
$\epsilon_K$ contributes to the asymmetry since the final states given in
$(\mathrm{a})$ and $(\mathrm{b})$
contain $K_L$ and $K_S$. We extracted the effect of $\epsilon_K$ and showed the contribution to an observable
in Eq.(\ref{2.13}). Formulation in this article is phase 
convention independent so that the coefficients in
Eqs.(\ref{2.13}) are invariant
under phase redefinition of quarks.\par
Furthermore, we discussed the conditions for the asymmetry to be a T-odd quantity. The two types of such a condition are
investigated; (1) B-meson state equivalence  and
(2) Vanishment of T-even part of overall normalizations' ratio of
the decay rates which form the asymmetry (denoted as $\Delta N_R^e=0$).
The condition (1) requires \cite{Applebaum:2013wxa} that the B-meson state which appears in the process in Fig.\ref{fig1}
should be equivalent to the B-meson which reveals in a process whose time direction
is totally opposite to Fig.\ref{fig1}.
The condition (2) is regarding to the difference of overall normalization of the rates.
We consider the event number asymmetry
so that one should take into account the condition (2)
unlike BaBar asymmetry \cite{Lees:2012uka}.
One found that the T-even part of $\Delta N_R$ yields
T-even contribution to the coefficients of the asymmetry.
We conclude that the coefficients of the event number asymmetry
become T-odd quantities when the conditions (1), (2) are simultaneously satisfied.
\section{Acknowledgment}
We would like to appreciate the organizers of FPCP2015 for the well-organized conference.\\
H. U. is supported by Hiroshima Univ. Alumni Association.

\end{document}